\documentclass{elsart}
\usepackage{amsfonts}
\usepackage{amssymb}
\usepackage{amsmath}
\usepackage{dcolumn}
\usepackage[all]{xy}
\usepackage[]{epsfig}
\begin{document}

\begin{frontmatter}

\title{Some Considerations About Podolsky-Axionic Electrodynamics}

\author{Patricio Gaete\thanksref{cile}}
\thanks[cile]{e-mail address: patricio.gaete@usm.cl}
\address{Departmento de F\'{\i}sica and Centro
Cient\'{\i}fico-Tecnol\'{o}gico de Valpara\'{\i}so, Universidad
T\'ecnica Federico Santa Mar\'{\i}a, Valpara\'{\i}so, Chile}

\begin{abstract}
For a Podolsky-axionic electrodynamics, we compute the interaction
potential within the structure of the gauge-invariant but
path-dependent variables formalism. The result is equivalent to that
of axionic electrodynamics from a new noncommutative approach, up to
first order in $\theta$.
\end{abstract}
\end{frontmatter}

\section{Introduction}

Higher order derivative theories have been discussed in the
literature by a large number of authors \cite{Bopp, Podolsky,
Schwed, Pais, Carlos, Cesar, Accioly, Scatena, Turcati,
Santos,Thirring}, mainly due to the possibility of obtaining finite
theories at short distances.  An illustrative example of such a
class of theories is the electrodynamics proposed by Podolsky
\cite{Podolsky}, i.e. the $U(1)$ gauge theory where a quadratic term
in the divergence of the field strength tensor is added to the free
Lagrangian of the $U(1)$ sector. As a result, this new theory is
endowed with interesting features such as a finite electron
self-energy and a regular point charge electric field at the origin.
In this context it may be recalled that field theories with higher
derivatives have also attracted considerable attention in connection
with supersymmetric \cite{West} and string theories \cite{Polyakov}.
Another interesting theory is the electrodynamics proposed by Lee
and Wick \cite{Lee, Wick}, i.e. the $U(1)$ gauge theory where a
dimensional-$6$ operator containing higher derivatives is added to
the free Lagrangian of the $U(1)$ sector. In this connection we call
attention to the fact that, following this proposal, recently a
great deal of attention has been devoted to the study  of
modifications to the Standard Model which stabilizes the Higgs mass
against quadratically divergent corrections leading to the Lee-Wick
Standard Model \cite{Grinstein}. More recently, by using a novel way
to formulate noncommutative quantum field theory (or quantum field
theory in the presence of a minimal length) \cite{Anais, Euro,
Piero} we have obtained an ultraviolet finite electrodynamics
\cite{Patricio}, where the cutoff is provided by the noncommutative
parameter $\theta$.

On the other hand, we also point out that recently considerable
attention has been paid to the formulation and experimental
consequences of extensions of the Standard Model such as axion-like
particles. Mention should be made, at this point, to axionic
electrodynamics which experiences mass generation due to the
breaking of rotational invariance induced by a classical background
configuration of the gauge field strength \cite{Stefano}. Interestingly
enough, we mention that axionic electrodynamics leads to a confining
potential in the presence of a nontrivial constant expectation
values for the gauge field strength $F_{\mu\nu}$ \cite{GaeteGue}.

In this perspective, and given the recent interest in higher order
derivative theories, it is instructive to further explore the
physical consequences presented by this class of theories.
Specifically, in this work we will focus attention on the impact of
these higher-order terms on physical observables, in particular the
static potential between two charges, using the gauge-invariant but
path-dependent variables formalism, which provides a
physically-based alternative to the Wilson loop approach \cite{GaeteSch, GaeteEuro1, GaeteEuro2, GaeteEuro3}.
In fact, our analysis leads to an ultraviolet finite static
potential for axionic electrodynamics which is the sum of a
Yukawa-type and a linear potential, leading to the confinement of
static charges. Incidentally, the above static potential profile is
equivalent to that of our noncommutative axionic electrodynamics up
to first order in $\theta$. In this way we establish a connection
between noncommutative and Podolsky electrodynamics, Although a
preliminary analysis about these issues has appeared before
\cite{Patricio, Jose}, we think is of value to clarify them because, in our
view, have not been properly emphasized. In addition, the above
connections are of interest from the point of view of providing
unifications among diverse models. The paper is organized as
follows. In the subsequent section, we present a brief review on
Podolsky electrodynamics, for which we compute the static potential
by two distinct methods. We show their equivalence and we set up the
baxkground for the Section III. In Section III, we consider the
Podolsky-axionic electrodynamics and we show that a confining
potential is achieved in the regime the axionic degree of freedom
decouples. A summary of our work is the subject of our final
section.

\section{Brief review on Podolsky electrodynamics}

We now reexamine the interaction energy between static pointlike
sources for Podolsky electrodynamics, through two different methods.
The first approach is based on the path-integral approach, whereas
the second one makes use of the gauge-invariant but path-dependent
variables formalism. The initial point of our analysis is the
Lagrangian density:
\begin{equation}
{\cal L} =  - \frac{1}{4}F_{\mu \nu } F^{\mu \nu }  + \frac{{l^2 }}
{2}\left( {\partial _\alpha  F^{\beta \alpha } } \right)^2  - A_\mu
J^\mu,  \label{Pod05a}
\end{equation}
which can also be written as
\begin{equation}
{\cal L} =  - \frac{1}{4}F_{\mu \nu } \left( {1 + l^2 \Delta }
\right) F^{\mu \nu } - A_\mu J^\mu,\label{Pod05b}
\end{equation}
where $l$ is a constant with dimension (-$1$) in mass units,
$\Delta\equiv \partial _\mu  \partial ^\mu$ and $J_\mu$ is an
external current. One immediately sees that the Lagrangian Eq.
(\ref{Pod05a}) (or (\ref{Pod05b})) contains second order time
derivatives of the potentials. This point motivates us to study the
role of higher order derivatives on a physical observable.

Let us start off our considerations by computing the interaction
energy between static pointlike sources for Podolsky electrodynamics
via a path-integral approach. For this purpose, we begin by writing
down the functional generator of the Green's functions, that is,
\begin{equation}
Z\left[ J \right] = \exp \left( { - \frac{i}{2}\int {d^4 xd^4 y}
J^\mu \left( x \right)D_{\mu \nu } \left( {x,y} \right)J^\nu  \left(
y \right)} \right), \label{Pod10}
\end{equation}
where $D_{\mu \nu } \left( {x,y} \right) = \int {\frac{{d^4
k}}{{\left( {2\pi } \right)^4 }}} D_{\mu \nu } \left( k \right)e^{ -
ikx}$ is the propagator in the Feynman gauge. In this case, the
corresponding propagator is given by
\begin{equation}
D_{\mu \nu } \left( k \right) =  - \frac{1}{{k^2 \left( {1 - l^2 k^2
} \right)}}\left\{ {\eta _{\mu \nu }  - \left[ {1 - \left( {1 - l^2
k^2 } \right)} \right]\frac{{k_\mu  k_\nu  }}{{k^2 }}} \right\}.
\label{Pod15}
\end{equation}
By means of expression $Z = e^{iW\left[ J \right]}$, and employing
Eq. (\ref{Pod10}), $W\left[ J \right]$ takes the form
\begin{equation}
W\left[ J \right] =  - \frac{1}{2}\int {\frac{{d^4 k}}{{\left( {2\pi
} \right)^4 }}} J_\mu ^ *  \left( k \right)\left[ { - \frac{1}{{k^2
\left( {1 - l^2 k^2 } \right)}}\eta ^{\mu \nu }  + \frac{{\left[ {1
- \left( {1 - l^2 k^2 } \right)} \right]}}{{k^2 \left( {1 - l^2 k^2
} \right)}}\frac{{k^\mu  k^\nu  }}{{k^2 }}} \right]J_\nu  \left( k
\right). \label{Pod20}
\end{equation}
Since the current $J^\mu (k)$ is conserved, expression (\ref{Pod20})
then becomes
\begin{equation}
W\left[ J \right] = \frac{1}{2}\int {\frac{{d^4 k}}{{\left( {2\pi }
\right)^4 }}J_\mu ^ *  } \left( k \right)\frac{1}{{k^2 \left( {1 -
l^2 k^2 } \right)}}J^\mu  \left( k \right). \label{Pod25}
\end{equation}
Next, for $J_\mu  \left( x \right) = \left[ {Q\delta ^{\left( 3
\right)} \left( {{\bf x} - {\bf x}^{\left( 1 \right)} } \right) + Q^
\prime  \delta ^{\left( 3 \right)} \left( {{\bf x} - {\bf x}^{\left(
2 \right)} } \right)} \right]\delta _\mu ^0$, and using standard
functional techniques \cite{Zee, Barone}, we obtain that the interaction
energy of the system is given by
 \begin{equation}
U(r) = QQ^ \prime  \int {\frac{{d^3 k}}{{\left( {2\pi } \right)^3
}}} \frac{{m^2 }}{{{\bf k}^2 \left( {{\bf k}^2  + m^2 }
\right)}}e^{i{\bf k} \cdot {\bf r}} , \label{Pod30}
\end{equation}
where $ {\bf r} \equiv {\bf x}^{\left( 1 \right)}  - {\bf x}^{\left(
2 \right)}$ and $m^2  = {\raise0.7ex\hbox{$1$} \!\mathord{\left/
{\vphantom {1 {l^2
}}}\right.\kern\nulldelimiterspace}\!\lower0.7ex\hbox{${l^2 }$}}$.
This, together with $Q^\prime=-Q$, yields finally
\begin{equation}
U(r) = - \frac{{Q^2 }}{{4\pi r}}\left( 1 - e^{ - mr} \right) ,
\label{Pod35}
\end{equation}
with $r = |{\bf r}|$. From this expression it should be clear that
the interaction energy is regular at the origin, in contrast to the
usual Maxwell theory. In this respect the above result clearly shows
the key role played by the "regularized propagator" in Eq.
(\ref{Pod25}) .

We shall now calculate the static potential  using this time a
gauge-invariant but path-dependent variables formalism along the
lines of Refs. \cite{GaeteSch, GaeteEuro1, GaeteEuro2, GaeteEuro3}.
To this end, we will compute the
expectation value of the energy operator $H$ in the physical state
$|\Phi\rangle$ describing the sources, which we will denote by
${\langle H\rangle}_\Phi$. In such a case, to obtain the
corresponding Hamiltonian, we must carry out the quantization of the
theory. Before going into details, we recall that the system
described by (\ref{Pod05a}) contains second order derivatives, hence
to construct the Hamiltonian one must consider the velocities as
independent canonical  variables. Thus, the phase-space coordinate
for the theory under consideration is given by $\left( {A_\mu  ,\Pi
^\nu  } \right) \oplus \left( {\dot A_\mu  ,\Pi^{\left( 1 \right)\nu
} } \right)$, where $\Pi ^{\left( 1 \right)\nu }$ is the canonical
momentum conjugate to $ \dot A_\mu$. This consideration implies that
the canonical Hamiltonian $H_{C}$ takes the form
\begin{equation}
H_C  = \int {d^3 x\left( {\Pi _\mu  \dot A^\mu   + \Pi _\mu ^{\left(
1 \right)} \ddot A^\mu   - {\cal L}} \right)} . \label{Pod45}
\end{equation}
In this case, the momenta are given by:
\begin{equation}
\Pi _\mu   = F_{\mu 0}  - l^2 \left[ {\partial ^\nu  \dot F_{\nu \mu
} + \eta _{\mu 0} \partial ^i \partial ^\nu  F_{i\nu }  + \partial
_\mu
\partial ^\nu  F_{0\nu } } \right], \label{Pod50a}
\end{equation}
and
\begin{equation}
\Pi _\mu ^{\left( 1 \right)}  =  - l^2 \left[ {\partial ^\nu  F_{\mu
\nu } + \eta _{\mu 0} \partial ^\nu  F_{\nu 0} } \right].
\label{Pod50b}
\end{equation}
It is not hard to check that, in the electrostatic case ($\dot{\bf
E}=0$ and ${\bf B}=0$), $\Pi _0^{\left( 1 \right)}  = 0$, $\Pi
_i^{\left( 1 \right)}  = 0$ and  $\Pi _0  = 0$. Therefore, in the
electrostatic case under consideration, the canonical Hamiltonian is
computed via the standard Legendre transformation. Accordingly, the
canonical Hamiltonian reduces to
\begin{equation}
H_C  = \int {d^3 x} \left\{ {\Pi ^i \partial _i A_0  -
\frac{1}{2}\Pi ^i \left( {1 - \frac{\Delta }{{\Delta  + m^2 }}}
\right)\Pi _i  + \frac{1}{4}F_{ij} \frac{{\left( {\Delta  + m^2 }
\right)}}{{m^2 }}F^{ij} } \right\}. \label{Pod60}
\end{equation}
Notice that, for notational convenience, we have maintained $\Delta$
in (\ref{Pod60}). As already explained, in the electrostatic case
under consideration $\Delta$ can be replaced by $- \nabla ^2$,
without any problem. Next, requiring the primary constraint $\Pi_0$
to be stationary,
leads to the secondary constraint, $\Gamma_1 \left(x \right) \equiv
\partial _i \Pi ^i=0$. It is straightforward to check that there are
no further constraints in the theory. Consequently, the extended
Hamiltonian that generates translations in time then reads
$H = H_C + \int {d^3 }x\left( {c_0 \left( x \right)\Pi _0 \left( x \right)
+ c_1 \left( x\right)\Gamma _1 \left( x \right)} \right)$. Here $c_0
\left( x\right)$ and $c_1 \left( x \right)$ are arbitrary Lagrange
multipliers. Moreover, it follows from this Hamiltonian that
$\dot{A}_0 \left( x \right)= \left[ {A_0 \left( x \right),H} \right]
= c_0 \left( x \right)$, which is an arbitrary function. Since
$\Pi^0 = 0$ always, neither $ A^0 $ nor $ \Pi^0 $ are of interest in
describing the system and may be discarded from the theory. The
Hamiltonian is then
\begin{equation}
H  = \int {d^3 x} \left\{ {c(x) \partial _i \Pi ^i- \frac{1}{2}\Pi
^i \left( {1 - \frac{\Delta }{{\Delta  + m^2 }}} \right)\Pi _i  +
\frac{1}{4}F_{ij} \frac{{\left( {\Delta  + m^2 } \right)}}{{m^2
}}F^{ij} } \right\},  \label{Pod65}
\end{equation}
where $c(x) = c_1 (x) - A_0 (x)$.

We can at this stage impose a gauge condition, so that in
conjunction with the constraint $\Pi^0=0$, it is rendered into a
second class set. A particularly convenient choice is
\begin{equation}
\Gamma _2 \left( x \right) \equiv \int\limits_{C_{\xi x} } {dz^\nu }
A_\nu \left( z \right) \equiv \int\limits_0^1 {d\lambda x^i } A_i
\left( {\lambda x} \right) = 0, \label{Pod70}
\end{equation}
where  $\lambda$ $(0\leq \lambda\leq1)$ is the parameter describing
the spacelike straight path $ x^i = \xi ^i  + \lambda \left( {x -
\xi } \right)^i $, and $ \xi $ is a fixed point (reference point).
There is no essential loss of generality if we restrict our
considerations to $ \xi ^i=0 $. The choice (\ref{Pod70}) leads to
the Poincar\'e gauge  \cite{Gaete99, Gaete97}. The Dirac brackets
can now be
determined and we simply note the only nontrivial Dirac bracket
involving the field variables,
\begin{equation}
\left\{ {A_i \left( x \right),\Pi ^j \left( y \right)} \right\}^ *
=\delta{ _i^j} \delta ^{\left( 3 \right)} \left( {x - y} \right) -
\partial _i^x \int\limits_0^1 {d\lambda x^j } \delta ^{\left( 3
\right)} \left( {\lambda x - y} \right). \label{Pod75}
\end{equation}

Our next task is to compute the interaction energy. As mentioned
before, to do that we need to calculate the expectation value of the
energy operator $H$ in the physical state $|\Phi\rangle$. Following
Dirac \cite{Dirac},  we write the physical state $|\Phi\rangle$  as
\begin{equation}
\left| \Phi  \right\rangle \equiv \left| {\overline \Psi  \left( \bf
y \right)\Psi \left( {\bf y}\prime \right)} \right\rangle =
\overline \psi \left( \bf y \right)\exp \left( {iq\int\limits_{{\bf
y}\prime}^{\bf y} {dz^i } A_i \left( z \right)} \right)\psi
\left({\bf y}\prime \right)\left| 0 \right\rangle, \label{Pod80}
\end{equation}
where the line integral is along a spacelike path on a fixed time
slice, $q$ is the fermionic charge, and $\left| 0 \right\rangle$ is
the physical vacuum state.

Taking  the above Hamiltonian structure into account, we see that
\begin{equation}
\Pi _i \left( x \right)\left| {\overline \Psi  \left( \bf y
\right)\Psi \left( {{\bf y}^ \prime  } \right)} \right\rangle  =
\overline \Psi  \left( \bf y \right)\Psi \left( {{\bf y}^ \prime }
\right)\Pi _i \left( x \right)\left| 0 \right\rangle +  q\int_ {\bf
y}^{{\bf y}^ \prime  } {dz_i \delta ^{\left( 3 \right)} \left( {\bf
z - \bf x} \right)} \left| \Phi \right\rangle. \nonumber\\
\label{Pod85}
\end{equation}
Therefore, $\left\langle H \right\rangle _\Phi$ can be written  as
\begin{equation}
\left\langle H \right\rangle _\Phi   = \left\langle H \right\rangle
_0 + \left\langle H \right\rangle _\Phi ^{\left( 1 \right)}  +
\left\langle H \right\rangle _\Phi ^{\left( 2 \right)},
\label{Pod90}
\end{equation}
where $\left\langle H \right\rangle _0  = \left\langle 0
\right|H\left| 0 \right\rangle$. The $\left\langle H \right\rangle
 _\Phi ^{\left( 1 \right)}$ and $\left\langle H \right\rangle
 _\Phi ^{\left( 2 \right)}$
terms are given by
\begin{equation}
\left\langle H \right\rangle _\Phi ^{\left( 1 \right)}  =  -
\frac{1}{2} \left\langle \Phi  \right|\int {d^3 x} \Pi _i \Pi ^i
\left| \Phi  \right\rangle, \label{Pod95a}
\end{equation}
and
\begin{equation}
\left\langle H \right\rangle _\Phi ^{\left( 2 \right)}  =
\frac{{1}}{2} \left\langle \Phi  \right|\int {d^3 x} \Pi _i
\frac{\nabla ^2} {{\left( {\nabla ^2  - m^2 } \right)}}\Pi ^i \left|
\Phi \right\rangle.  \label{Pod95b}
\end{equation}
Using Eq. (\ref{Pod85}), and following our earlier procedure
\cite{GaeteSch, GaeteEuro1, GaeteEuro2, GaeteEuro3}, we see that the potential for two opposite charges,
located at ${\bf y}$ and ${\bf y}'$, takes the form
\begin{equation}
U(r) =  - \frac{{q^2 }}{{4\pi r }}\left( 1 - e^{ - mr}  \right),
\label{Pod96}
\end{equation}
where $\left|{\bf y} - {\bf y}'\right|=r$. It is worth noting that
these approaches, despite being completely different, lead to the
same result which seems to indicate that they are equivalent term by
term. It should, however, be noted here that the central difference
between the above analysis and that leading to Eq. (\ref{Pod35})
rests in the fact that the potential (\ref{Pod96}) is directly
recovered from the constraints structure of the theory we have
discussed.

One can now further observe that there is an alternative but
equivalent way of obtaining the result \cite{GaeteSch, Gaete99, Gaete97}, which highlights certain distinctive features of our methodology. In order
to illustrate the discussion, we start by observing that:
\begin{equation}
U \equiv q\left( {{\cal A}_0 \left( {\bf 0} \right) - {\cal A}_0
\left( {\bf y} \right)} \right), \label{Pod97}
\end{equation}
where the physical scalar potential is given by
\begin{equation}
{\cal A}_0 \left( {x^0 ,{\bf x}} \right) = \int_0^1 {d\lambda } x^i
E_i \left( {\lambda {\bf x}} \right), \label{Pod100}
\end{equation}
with $i=1,2,3$. This follows from the vector gauge-invariant field
expression:
\begin{equation}
{\cal A}_\mu  \left( x \right) \equiv A_\mu  \left( x \right) +
\partial _\mu  \left( { - \int_\xi ^x {dz^\mu  } A_\mu  \left( z
\right)} \right), \label{Pod105}
\end{equation}
where, as in Eq. (\ref{Pod80}), the line integral is along a
spacelike path from the point $\xi$ to $x$, on a fixed slice time.
The gauge-invariant variables (\ref{Pod105}) commute with the sole
first constraint (Gauss' law), confirming in this way that these
fields are physical variables. Note that Gauss' law for the present
theory reads $\partial _i \Pi ^i  = J^0$, where we have included the
external current $J^0$ to represent the presence of two opposite
charges. For $J^0 \left( {t,{\bf x}} \right) = q\delta ^{\left( 3
\right)} \left( {\bf x} \right)$ the electric field is given by
\begin{equation}
E^i  = q\partial ^i \left( {G\left( {\bf x} \right) - G^ \prime
\left( {\bf x} \right)} \right), \label{Pod110}
\end{equation}
where $G\left( {\bf x} \right) = \frac{1}{{4\pi }}\frac{1}{{|{\bf
x}|}}$ and $G'\left( {\bf x} \right) = \frac{e^{ - m  \left| {\bf x}
\right|}}{4\pi \left| {\bf x} \right|}$ are the Green functions in
three space dimensions. Finally, replacing this result in
(\ref{Pod100}) and using (\ref{Pod97}), we reobtain Eqs.
(\ref{Pod35}) and (\ref{Pod96}), i.e.,
\begin{equation}
U(r) =  - \frac{{q^2 }}{{4\pi r}}\left( 1 - e^{ - mr}  \right).
\label{Pod115}
\end{equation}
Notice that the procedure leading to Eq. (\ref{Pod35}) , compared
with the above one, involves the gauge field propagator, which leads
us to conclude that the contributions of the propagator are properly
captured in the gauge-invariant variables formalism. This concludes
our considerations about Podolsky electrodynamics.

\section{Podolsky-Axionic electrodynamics}

As already stated, our next undertaking is to use the ideas of the
previous section in order to consider Podolsky-axionic
electrodynamics. In such a case the Lagrangian density reads
\begin{equation}
\mathcal{L} =  - \frac{1}{4}F_{\mu \nu } \left( {1 + l^2 \Delta }
\right)F^{\mu \nu } + \frac{1}{2} \left( {\partial _\mu  \varphi }
\right)^2  - \frac{1} {2}\mu ^2 \varphi ^2  + \frac{\lambda }{4}\varphi
\widetilde F^{\mu \nu } F_{\mu \nu}. \label{Pod120}
\end{equation}

Before we proceed to work out explicitly the energy, let us commence
our considerations with a short presentation of previous results
stemming from the gauge-invariant formalism \cite{GaeteSch, GaeteEuro1, GaeteEuro2, GaeteEuro3}. For this
purpose, we carry out the integration over the $\varphi$-field.
Furthermore, as was explained in \cite{GaeteGue}, by considering static
scalar fields we may replace  $\Delta\varphi = - \nabla ^2\varphi$.
In this case, the effective theory takes the form
 \begin{equation}
\mathcal{L} =  - \frac{1}{4}F_{\mu \nu } \left( {1 + l^2 \Delta }
\right)F^{\mu \nu }    - \frac{{\lambda ^2 }} {{32}}\left( {
\widetilde F_{\mu \nu } F^{\mu \nu } } \right)\frac{1} {{\nabla ^2 -
\mu ^2 }}\left( { \widetilde F_{\alpha \beta } F^{\alpha \beta }
} \right). \label{Pod125}
\end{equation}
Following our earlier discussion, after splitting $F_{\mu \nu }$ in
the sum of a classical background, $\langle F_{\mu \nu }\rangle$,
and a small fluctuation, $f_{\mu \nu }$, the Lagrangian
(\ref{Pod125}) up to quadratic terms in the fluctuations then
becomes
\begin{equation}
\mathcal{L} =  - \frac{1}{4}f_{\mu \nu } \left( {1 + l^2 \Delta }
\right)f^{\mu \nu } - \frac{{\lambda ^2 }} {{32}}v^{\mu \nu } f_{\mu
\nu } \frac{1}{{\nabla ^2  - \mu ^2 }}v^{\lambda \rho } f_{\lambda \rho
}, \label{Pod130}
\end{equation}
where $f_{\mu\nu} = \partial_\mu a_\nu - \partial_\nu a_\mu$,
$a_{\mu}$ stands for the fluctuation, and $\varepsilon ^{\mu \nu
\alpha \beta } \left\langle {F_{\alpha \beta } } \right\rangle
\equiv v^{\mu \nu }$ and $\varepsilon ^{\rho\lambda\gamma\delta }
\left\langle {F_{\gamma\delta } } \right\rangle \equiv
v^{\rho\lambda }$.

After having obtained the general effective theory, we now turn our
attention to the calculation of the interaction energy in the
$v^{0i}  \ne 0$ and $v^{ij} = 0$ case (referred to as the electric
one in what follows). In such a case, the Lagrangian (\ref{Pod130})
reduces to
\begin{equation}
\mathcal{L} =  - \frac{1}{4}f_{\mu \nu } \left( {1 + l^2 \Delta }
\right)f^{\mu \nu } - \frac{{\lambda ^2 }} {{32}}v^{0i} f_{0i}
\frac{1}{{\nabla ^2  - \mu ^2}}v^{0k} f_{0k}. \label{Pod135}
\end{equation}

It is now once again straightforward to apply the formalism
discussed in the preceding section. Therefore the canonical
Hamiltonian is
\begin{equation}
H_C  = \int {d^3 x} \left\{ {\Pi _i \partial ^i a^0  +
\frac{1}{2}\Pi ^i \left[ {\frac{{\nabla ^2  - \mu ^2 }}{{\left( {1 +
l^2 \Delta } \right)\left( {\nabla ^2  - \mu ^2 } \right) - w^2 }}}
\right]\Pi ^i + \frac{1}{2}{\bf B}^2 } \right\}, \label{Pod140}
\end{equation}
where $w^2  = \frac{{\lambda ^2 }}{4}{\cal B}^2$. Here $\bf B$  and
$\cal B$    represent the magnetic field fluctuation and external
(background) magnetic field.

By means of expression (\ref{Pod85}) we write the expectation value
as
\begin{equation}
\left\langle H \right\rangle _\Phi   = \left\langle H \right\rangle
_0 + \left\langle H \right\rangle _\Phi ^{\left( 1 \right)}  +
\left\langle H \right\rangle _\Phi ^{\left( 2 \right)},
\label{Pod145}
\end{equation}
with  $\left\langle H \right\rangle _0  = \left\langle 0
\right|H\left| 0
 \right\rangle$,
while the terms $\left\langle H \right\rangle _\Phi ^{\left( 1
\right)}$ and $\left\langle H \right\rangle _\Phi ^{\left( 2
\right)}$ are given by
\begin{eqnarray}
\left\langle H \right\rangle _\Phi ^{\left( 1 \right)}  &=&  -
\frac{1} {2}\frac{1}{{\sqrt {1 - 2l^2 \left( {\mu ^2  + 2w^2 } \right)
+ \mu ^4 l^4 } }}  \nonumber\\
&\times& \left\langle \Phi  \right|\int {d^3 x} \Pi _i \left\{
{\frac{{\nabla ^2 }} {{(\nabla ^2  - M_2^2 )}} - \frac{{\nabla ^2
}}{{(\nabla ^2  - M_1^2 )}}} \right\}\Pi ^i \left| \Phi
\right\rangle, \label{Pod150}
\end{eqnarray}
and
\begin{eqnarray}
\left\langle H \right\rangle _\Phi ^{\left( 2 \right)}  &=&
\frac{1}{2} \frac{1}{{\sqrt {1 - 2l^2 \left( {\mu ^2  + 2w^2 } \right)
+ \mu ^4 l^4 } }} \nonumber\\
&\times& \left\langle \Phi  \right|\int {d^3 x} \Pi _i \left\{
{\frac{1}{{ (\nabla ^2  - M_1^2 )}} - \frac{1}{{(\nabla ^2  - M_2^2
)}}} \right\} \Pi ^i \left| \Phi  \right\rangle. \label{Pod155}
\end{eqnarray}
Here $M_1^2  = \frac{1}{{2l^2 }}\left[ {\left( {1 + \mu ^2 l^2 }
\right) + \sqrt {1 - 2l^2 \left( {\mu ^2  + 2w^2 } \right) + \mu ^4 l^4 }
} \right]$ and \\ $ M_2^2  = \frac{1}{{2l^2 }}\left[ {\left( {1 + \mu ^2
l^2 } \right) -
\sqrt {1 - 2l^2 \left( {\mu ^2  + 2w^2 } \right) + \mu ^4 l^4 } } \right]$,
where $|1 - \mu ^2 l^2 | > 2\lambda \omega$, which ensures
$M_1^2  > 0$ and $M_2^2  > 0$.

Accordingly, the potential for a pair of point-like opposite charges
q located at ${\bf 0}$ and ${\bf L}$ takes the form
\begin{eqnarray}
U &=&  - \frac{{q^2 }}{{4\pi }}\frac{1}{{\sqrt {1 - 2l^2 \left( {\mu ^2 +
2w^2 } \right) + \mu ^4 l^4 } }}\left[ {\frac{{e^{ - M_2 L} }}{L} -
\frac{{e^{ - M_1 L} }}{L}} \right] \nonumber\\
&+& \frac{{q^2 }}{{8\pi
}}\frac{{\mu ^2 }}{{\sqrt {1 - 2l^2 \left( {\mu ^2  + 2w^2 } \right) + \mu ^4
l^4 } }}\ln \left( {\frac{{M_2^2 }}{{M_1^2 }}} \right) L.
\label{Pod160}
\end{eqnarray}
Expression (\ref{Pod160})  immediately shows that the effect of
including higher order derivative terms is an ultraviolet finite
static potential, which is the sum of a Yukawa and a linear
potential, leading to the confinement of static charges. Another
interesting finding is the presence of a finite string  tension in
Eq. (\ref{Pod160}). Evidently, this improves the analysis as
compared to our previous studies \cite{GaeteGue, GaeteEuro4},
where an ultraviolet
cutoff has been introduced by hand.

\section{Final Remarks}

Let us summarize our work. Using the gauge-invariant but
path-dependent formalism, we have computed the static potential for
Podolsky-axionic electrodynamics. Interestingly, we have obtained an
ultraviolet finite static potential, which is the sum of a
Yukawa-type and a linear potential, leading to the confinement of
static charges. As already expressed,  the above static potential
profile is equivalent to that of our noncommutative axionic
electrodynamics up to first order in $\theta$. In this way we have
provided a new connection between effective models. Accordingly, the
benefit of considering the present framework is to provide
unifications among different models, as well as exploiting the
equivalence in explicit calculations, as we have seen in the course
of the present discussion. Finally, an explicit expression of the
effective potential between two static charges could be of interest
for searching for bounds of the constant $l$.

\section{Acknowledgments}

It is a pleasure to thank J. A. Hela\"{y}el-Neto for collaboration
and useful discussions. This work was supported in part by Fondecyt
(Chile) grant 1080260.


\begin{thebibliography}{}
\bibitem{Bopp} F. Bopp,  Ann. Phys. (Paris) {\bf 38}, 345 (1940).
\bibitem{Podolsky} B. Podolsky,  Phys. Rev. {\bf 62}, 68 (1942).
\bibitem{Schwed} B. Podolsky and P. Schwed, Rev. Mod. Phys. {\bf 20},
4 (1948).
\bibitem{Pais} A. Pais and G. E. Uhlenbeck, Phys. Rev. {\bf 79}, 145 (1950).
\bibitem{Carlos} C. A. P. Galv\~ao and B. M. Pimentel, Can. J. Phys. {\bf 66},
460 (1988).
\bibitem{Cesar} J. Barcelos-Neto, C. A. P. Galv\~ao and C. P. Natividade,
Z. Phys. C {\bf 52}, 559 (1991).
\bibitem{Accioly} A. Accioly and M. Dias, Phys. Rev. D {\bf 70}, 107705 (2004).
\bibitem{Scatena} A. Accioly and E. Scatena, Mod. Phys. Lett. A {\bf 25},
269 (2010).
\bibitem{Turcati} A. Accioly, P. Gaete, J. Helay\"el-Neto, E. Scatena and
R. Turcati, Mod. Phys. Lett. A {\bf 26}, 1985 (2011).
\bibitem{Santos} R. B. Santos, Mod. Phys. Lett. A {\bf 26},  1909 (2011).
\bibitem{Thirring} H. Narnhofer and W. Thirring, Phys. Lett. B {\bf
76}, 428 (1978).
\bibitem{West} P. West, Nucl. Phys. B {\bf 268}, 113 (1986).
\bibitem{Polyakov} A. Polyakov, Nucl. Phys. B {\bf 268}, 406 (1986).
\bibitem{Lee} T. Lee and G. Wick, Nucl. Phys. B {\bf 9}, 209 (1969).
\bibitem{Wick} T. Lee and G. Wick, Phys. Rev. D {\bf 2}, 1033 (1970).
\bibitem{Grinstein} B. Grinstein, D. O'Connell, and M. Wise,  Phys. Rev. D
{\bf 77}, 025012 (2008).
\bibitem{Anais}  A. Smailagic and E. Spallucci, J. Phys. A {\bf 36},
L517 (2003).
\bibitem{Euro}  A. Smailagic and E. Spallucci, J. Phys. A {\bf 36},
L467 (2003).
\bibitem{Piero} P. Nicolini, A. Smailagic and E. Spallucci, Phys. Lett.
B {\bf 632}, 547 (2006).
\bibitem{Patricio}  P. Gaete and  E. Spallucci, arXiv: 1102.3777 [hep-th].
\bibitem{Stefano} S. Ansoldi, E. I. Guendelman and  E. Spallucci, J. High.
Energy Phys. {\bf 09}, 044 (2003).
\bibitem{GaeteGue} P. Gaete and E. I. Guendelman, Mod. Phys. Lett. A {\bf 20}, 319 (2005).
\bibitem{GaeteSch} P. Gaete and I. Schmidt, Phys. Rev. D {\bf61}, 125002 (2000).
\bibitem{GaeteEuro1} P. Gaete and E. Spallucci, Phys. Rev. D {\bf 77}, 027702 (2008).
\bibitem{GaeteEuro2}  P. Gaete and E. Spallucci, J. Phys. A: Math. Theor. {\bf41}, 185401 (2008).
\bibitem{GaeteEuro3} P. Gaete and E. Spallucci, Phys. Lett. B {\bf675}, 145 (2009).
\bibitem{Jose} A. Accioly, P. Gaete, J. Helay\"el-Neto, E. Scatena and
R. Turcati, arXiv: 1012.1045 [hep-th].
\bibitem{Zee}  A. Zee, {\it Quantum Field Theory in a Nutshell}
(Princeton University Press, Princeton, NJ, 2003).
\bibitem{Barone} F. A. Barone and G. Fores-Hidalgo, Phys. Rev. D {\bf 78},
125003 (2008).
\bibitem{Gaete99} P. Gaete, Phys. Rev. D {\bf59}, 127702 (1999).
\bibitem{Gaete97}  P. Gaete, Z. Phys. C {\bf76}, 355 (1997).
\bibitem{Dirac}  P. Dirac, Can. J. Phys. {\bf33}, 650 (1955).
\bibitem{GaeteEuro4}  P. Gaete and E. Spallucci, J. Phys. A {\bf 41}, 185401 (2008).
\end{thebibliography}
\end{document}